\documentclass[epj]{svjour}

\usepackage[pdftex]{graphicx}
\usepackage{amsmath,amsfonts,amssymb}

\begin{document}

\title{Truncated L\'evy Random Walks and Generalized Cauchy Processes}

\author{Ihor Lubashevsky\thanks{\,\email{i-lubash@u-aizu.ac.jp}}}

\institute{
\mbox{University of Aizu, Ikki-machi, Aizu-Wakamatsu City, Fukushima 965-8560, Japan}
}
\date{Received: date / Revised version: date}
%
\abstract{A continuous Markovian model for truncated L\'evy random walks is proposed. It generalizes the approach developed previously by Lubashevsky et al. Phys. Rev. E \textbf{79}, 011110 (2009); \textbf{80}, 031148 (2009), Eur. Phys. J. B \textbf{78}, 207 (2010) allowing for nonlinear friction in wondering particle motion  and saturation of the noise intensity depending on the particle velocity. Both the effects have own reason to be considered and individually give rise to truncated L\'evy random walks as shown in the paper. The nonlinear Langevin equation governing the particle motion was solved numerically using an order 1.5 strong stochastic Runge-Kutta method and the obtained numerical data were employed to calculate the geometric mean of the particle displacement during a certain time interval and to construct its distribution function. It is demonstrated that the time dependence of the geometric mean comprises three fragments following one another as the time scale increases that can be categorized as the ballistic regime, the L\'evy type regime (superballistic, quasiballistic, or superdiffusive one), and the standard motion of Brownian particles. For the intermediate L\'evy type part the distribution of the particle displacement is found to be of the generalized Cauchy form with cutoff. Besides, the properties of the random walks at hand are shown to be determined mainly by a certain ratio of the friction coefficient and the noise intensity rather then their characteristics individually.
} 

\maketitle

\section{Introduction}

L\'evy random walks and L\'evy flights are met in a large variety of systems different in nature (see, e.g. \cite{Mandl,BG,Biol1,Zasl1,Klaft1,ChBook,RW1,RW7}). In the strict mathematical sense, such Markovian stochastic processes are characterized by divergence of the second moment of walker displacement $x(t)$ for any time scale $t$, i.e., $\left<[x(t)]^2\right> \to \infty$, which is caused by power-law asymptotics of the distribution function $P(x,t)$. For example, in the 1D case it is $P(x,t)\sim Dt/x^{1+\alpha}$ for $x\gg(Dt)^{1/\alpha}$, where $D$ is some constant and the exponent $0<\alpha<2$ specifies the superdiffusion law usually written in a \textit{symbolic} form as $\left<[x(t)]^2\right> \propto t^\beta$ with the exponent $\beta = 2/\alpha>1$.
Naturally, in the reality this divergence should be suppressed, for example, by the finite size of a system at hand. Thereby the given asymptotics can hold only within the frameworks of certain spatial and temporal scales bounded from below and above. The notion of truncated L\'evy random walks (flights) takes into account the existence of these boundaries.  To allow for such cutoff effects explicitly several approaches have been proposed. These include continuous-time random walks governed by the coupled spatial-temporal memory with finite moments \cite{CTRW1,CTRW2}, coupled continuous-time random walks with bounded variations in the velocity fluctuations \cite{CTRW3}, a direct cutoff in the L\'evy noise \cite{Cutoff1,Cutoff2}, L\'evy flights confined by external potentials \cite{Well1}, as well as  L\'evy flights damped by dissipative nonlinearity \cite{Checkin2005}.  

Papers~\cite{me1,me2,me3} developed a new approach to describing L\'evy random walks based on continuous Markovian nonlinear stochastic processes. For example, in the 1D case it deals with random motion of a particle along the $x$-axis whose velocity $v$ obeys the following stochastic differential equation
\begin{subequations}\label{Int:1}
\begin{equation}\label{Int:1a}
   \tau \frac{dv}{dt} = -\lambda v + \sqrt{2\tau(v_a^2 + v^2)}*\xi(t)
\end{equation}
written in the H\"anggi-Klimontovich form, which is indicated with the multiplication symbol $*$ in the product of the white noise $\xi(t)$ and the intensity $\sqrt{2\tau(v_a^2+v^2)}$ of the Langevin forces depending itself on the particle velocity $v$. Here $\tau$ is a certain ``microscopic'' time scale of the particle dynamics, $\lambda$ is a friction coefficient, and the parameter $v_a$ actually quantifies the intensity of ``additive'' noise. Indeed, a time patterns $\{v(t)\}$ with the same statistics can be generated using also the following stochastic differential equation 
\begin{equation}\label{Int:1b}
   \tau \frac{dv}{dt} = -\left[\lambda + \sqrt{2\tau} \xi_m(t) \right]*v + \sqrt{2\tau}v_a\xi_a(t)
\end{equation}
\end{subequations}
explicitly containing additive, $\xi_a(t)$, and multiplicative, $\xi_m(t)$, noise \cite{Konno2007}. It should be noted that equation~\eqref{Int:1b} has been much employed to study stochastic behavior of various nonequilibrium systems, in particular, lasers \cite{K1}, on-off intermittency \cite{Nak5}, economic activity \cite{Nak9}, passive scalar field advected by fluid \cite{Nak11}, etc. Besides, appealing to qualitative arguments and a numerical example  Sakaguchi \cite{Sak} demonstrated  that this equation can give rise to L\'evy flights in the space $x$ on time scales $t\gg \tau$ if the random variable $v$ is regarded as the velocity $v=dx/dt$ of a Brownian particle.

Random walks generated by model~\eqref{Int:1a}, on one hand, can be treated as continuous trajectories of particle motion and the corresponding probability density function $P(x,v,t)$ obeys the standard Fokker-Planck equation. It admits a conventional generalization taking into account possible medium heterogeneities as well as the existence of system boundaries via the appropriate boundary conditions. On the other hand, on time scales $t\gg\tau$ these random walks exhibit the characteristic properties, namely, the scaling law $\left<[x(t)]^2\right> \propto t^\beta$ and the power-law asymptotics of $P(x,v,t)$ as $x\to\infty$, that enable us to categorize them as L\'evy random walks. It has been proved analytically for the superdiffusive regime ($1<\beta<2$) \cite{me2} and verified numerically for the quasiballistic and superballistic regimes ($\beta\geq2$) \cite{me3}.  So the given approach seems to make it possible to attack the yet unsolved problem of the formulation of accurate boundary conditions for the fractional Fokker-Planck equations describing L\'evy processes in finite domains and heterogeneous media. 

The purpose of the present paper is to generalize the developed approach in such a way that explicitly allows for not only the small scale cutoff of L\'evy random walks determined actually by the parameter $\tau$ in the model at hand but also a large scale cutoff.

\subsubsection*{Generalized Cauchy processes}

We employ generalized Cauchy processes for constructing a model for truncated L\'evy random walks. There are several ways of introducing such stochastic processes starting from the Cauchy distribution 
\begin{equation*}
  P_C(z)\propto \frac1{(\overline{z}_t^2+ z^2)}
\end{equation*}
of a random variable $z$, where $\overline{z}_t$ is a certain parameter which, in principle, can depend on time $t$ if, e.g.,  $\{z(t)\}$ is some stochastic process. In particular, Lim and Li \cite{LL} following  Gneiting and Schlather \cite{GS2004} focused their attention on anomalous relaxation phenomena and introduced generalized Cauchy processes appealing to a power-law decay in correlation functions. For example, the Cauchy distribution meets this interpretation when $\overline{z}_t\propto t^{\beta/2}$, because in this case  $P_C(0)\propto t^{-\beta}$. Konno et al. \cite{Konno2007,Konno2011} related a generalization of Cauchy processes to a power-law asymptotics (maybe with some cutoff) of the distribution function $P(z)\propto |z|^{-\beta'}$ when the analyzed random variable $z$ takes large values,  $|z|\gg \overline{z}_t$. For the Cauchy distribution the equality $\beta'=2$ holds. Below the term of the generalized Cauchy processes will be understood in the latter sense.

Let us discuss a modification of the Langevin equation~\eqref{Int:1a} within the introduction of a nonlinear friction, i.e., a friction coefficient $\lambda(v)$ depending on the particle velocity $v$ as well as the intensity $G(v)$ of the Langevin random forces depending on the particle velocity $v$ in a more complex way then it is accepted in model~\eqref{Int:1a}. i.e., within the replacement
\begin{equation}\label{Int:2}
  \lambda\rightarrow\lambda(v)\quad\text{and}\quad \sqrt{v_a^2+v^2}\rightarrow G(v)\,.
\end{equation}
It should be pointed out that the introduction of nonlinear friction into a stochastic differential equation with additive L\'evy noise gives rise to truncated L\'evy flights \cite{Checkin2005} as well as the Langevin equation~\eqref{Int:1a} modified in the same way describes the generalized Cauchy process with a large scale cutoff \cite{Konno2011}.    

Giving consideration to each of the two replacements has its own reason. First, the velocity dependence of friction coefficient is widely used in describing nonlinear dissipation processes in nonequilibrium systems and for the systems where the intensity of dissipation grows with particle velocity the Ansatz    
\begin{equation}\label{Int:3}
    \lambda(v) = \lambda_0 + \lambda_2 v^2
\end{equation}
with the coefficients $\lambda_0,\lambda_2 >0 $ is a simple and rather natural model for these phenomena (see, e.g., \cite{KlimontovichPOS}). Moreover, in describing motion of biological objects models assuming $\lambda(v)\propto |v|^{\beta''}$ are also met (see, e.g., \cite{Lindner2007}). Appealing to Brownian motion governed by nonlinear friction and L\'evy noise \cite{Checkin2005} we may expect that a model similar to \eqref{Int:1a} with the friction coefficient~\eqref{Int:3} also generates truncated L\'evy random walks on large time scales.

\begin{figure}
 \begin{center}
   \includegraphics[width=0.8\columnwidth]{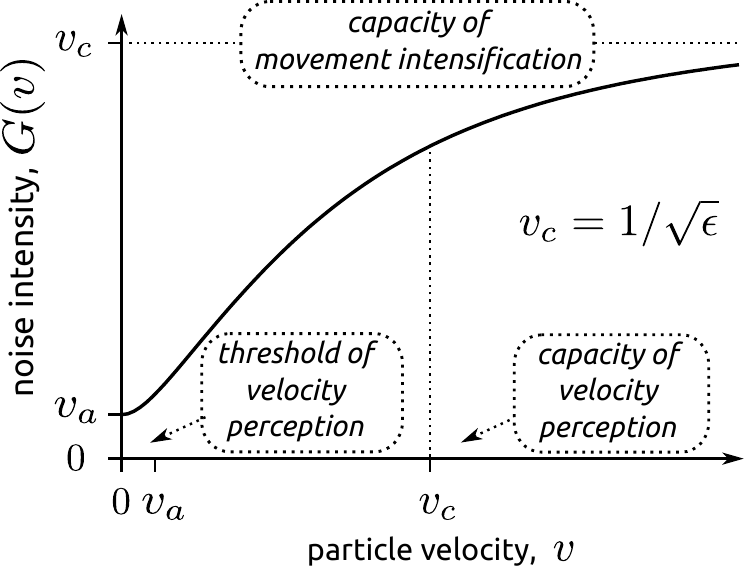}
\end{center}
\caption{The characteristic dependence of the noise intensity on the particle velocity that mimics the animal behavior in foraging. It was used in constructing Ansatz~\eqref{GonV}.}
\label{F1}
\end{figure}

Second, the cutoff effect can be also due to the intensity $G(v)$ of the Langevin random forces reaching its saturation as the velocity $v$ increases. By way of example, let us mention movement of animals in searching for food resources, mates, den sites, etc. It has been found out that when animals have no information about the targets the resulting movement patterns of many species are of fractal, i.e., scale-free structure at least within multiple scales and can be described in terms of L\'evy flights or truncated once  (see, e.g., \cite{RW1,RW7}). The complexity of the movement patterns is due to combination of various ``behavioral modes'' that change over time and, thus, this animal movement can be regarded as a composition of Gaussian random walks with the intensity changing in time \cite{Benhamou2007} (see also the following discussion \cite{BenhD1,BenhD2}). To quantify the intensity of the current behavioral mode the magnitude $|\mathbf{v}|$ of the animal velocity $\mathbf{v}$ can be used as a natural parameter that aggregates in itself the ultimate stimuli to the given behavior. So in mimicking the animal motion patterns based on stochastic differential equations similar to \eqref{Int:1a} the noise intensity $G(v)$ may be approximated by the Ansatz 
\begin{equation}\label{GonV}
    G(v) = \sqrt{\frac{v_a^2 +v^2}{1+ \epsilon v^2 }}\,.
\end{equation} 
It takes into account the low threshold in the animal perception of the optimal behavior as well as the animal bounded capacity of intensifying the search process quantified by the parameter $\epsilon$ such that $\epsilon v^2_a\ll 1$. Figure~\ref{F1} illustrates dependence~\eqref{GonV}.  

For the stochastic process governed by equation~\eqref{Int:1a} within replacements~\eqref{Int:2} the probability distribution function $P(v,t)$ obeys the following Fokker-Planck equation
\begin{equation}\label{FPE}
 \tau\frac{\partial P}{\partial t} = \frac{\partial }{\partial v}\left\{G^2(v)\frac{\partial P}{\partial v} + v \lambda(v) P \right\}\,.
\end{equation} 
Its steady state solution, i.e., the stationary distribution $P^\text{st}(v)$ is of the form
\begin{equation}\label{Pst}
 P^\text{st}(v) = C\exp\left[-\int\limits_0^v \frac{u\lambda(u)}{G^2(u)}\,du\right]\,,
\end{equation}
where $C$ is the normalization constant specified by the equality
\begin{equation}\label{C}
 \frac{1}{C} = 2 \int\limits_0^\infty \exp\left[-\int\limits_0^v \frac{u\lambda(u)}{G^2(u)}\,du\right]dv
\end{equation}
stemming directly from the normalization of the distribution function  $P^\text{st}(v)$ to unity.

In what follows we intend to analyze the two possible mechanisms of the cutoff individually. Namely, two alternative cases will be studied employing separately the former replacement of \eqref{Int:2} or the latter one. In order to have some common ``reference point'' in comparing the cutoff effects caused by these mechanisms we will consider the models characterized by the same stationary distribution function. In other words, the parameters $\lambda_0$, $\lambda_2$, and $\epsilon$ will be set equal to such values that the ratio $\lambda(v)/G^2(v)$ be the same function in both the cases. In addition, to single out the key points of the models at hand let us convert to the dimensionless variables, namely,
\begin{align}\label{Int:scaling}
  t &\rightarrow t \tau\,, &  x &\rightarrow (v_a\tau) x\,, & v&\rightarrow v_a v\,.
\end{align}
Below all the results will be presented using these dimensionless time $t$, spatial coordinate $x$, and velocity $v$.

\section{Model}

Random walks $\{x(t)\}$ of a wandering particle in the 1D space $x\in\mathbb{R}$ are under consideration. The dynamics of its velocity $v=dx/dt$ is assumed to be governed by the following  Langevin equation of the H\"anggi-Klimontovich type
\begin{equation}\label{eq:dv}
  \frac{dv}{dt}=-(\alpha + 1) v k(v)+\sqrt{2}g(v)*\xi(t)\,.
\end{equation}
Here $\xi(t)$ is the white Gaussian noise with the correlation function
\begin{equation}\label{eq:wgn}
  \left\langle \xi(t)\xi(t')\right\rangle =\delta(t-t')\,,
\end{equation}
the value $0<\alpha<2$ is a model parameter and $k(v)>0$, $g(v)>0$ are certain positive definite smooth functions of the argument $v^{2}$ such that
\begin{equation}\label{eq:fg0val}
k(0)  = 1 \quad \text{and} \quad g(0)  = 1\,.
\end{equation}
The numerical coefficient $\sqrt{2}$ has been introduced into \eqref{eq:dv} for the sake of convenience. In what follows the functions $k(v)>0$ and $g(v)>0$ will be referred to as the friction coefficient and the noise intensity, respectively, or just kinetic coefficients with respect to both of them.

When the kinetic coefficients are specified by the expressions
\begin{equation}\label{eq:fg0appr}
   k_0(v)=1\quad\text{and}\quad g_0(v) = \sqrt{1+v^2}\,,
\end{equation}
the stationary distribution function~\eqref{Pst} which in the given case takes the form  
\begin{equation}\label{LPst}
  P_0^\text{st}(v) = \frac{1}{B\left(\tfrac12,\tfrac\alpha2\right)\left(1+v^2\right)^{\tfrac{(1+\alpha)}{2}}}
\end{equation}
exhibits power-law asymptotics and its second moment diverges for $\alpha\leq 2$. Here $B(\cdot,\cdot)$ is the Beta function. Under such conditions stochastic motion of this particle can be regarded as L\'evy random walks on time scales $t\gg1$ \cite{me1,me2,me3}. Moreover, for $\alpha\in(0,1)$  the generated L\'evy random walks are of the superballistic type \cite{me3}, whereas for $\alpha\in(1,2)$ they are of the superdiffusive type \cite{me1,me2}. When $\alpha = 1$ the stationary distribution function of the particle velocity is of the Cauchy form and the corresponding random walks can be treated as quasi-ballistic ones \cite{me3}.

As stated in Introduction, in the present paper the cutoff effects are related to deviation of the kinetic coefficients $k(v)$ and $g(v)$ from the ideal dependences $k_0(v)$ and $g_0(v)$. To be specific we will make us of the following Ans\"atze 
\begin{align}\label{KGc}
  k_c(v) & = 1+\frac{v^2}{v_c^2}  &&\text{and} &  g_c(v)  & = \sqrt{\frac{1 +v^2}{1+ v^2/v_c^2 }}\,,
\end{align}
where the parameter $v_c\gg1$ is a certain critical value of the particle velocity when nonlinear effects responsible for the cutoff become essential.

Below we will consider in detail two cases, 
\begin{equation*}
 \text{case I}\Rightarrow
    \begin{cases}
          k(v) = k_c(v) \\
          g(v) = g_0(v)
    \end{cases}
 \!\!\!\text{and case II}\Rightarrow
    \begin{cases}
          k(v) = k_0(v) \\
          g(v) = g_c(v)
    \end{cases}
\end{equation*} 
which enables us, in particular, to elucidate the difference between the L\'evy type random works generated by the stochastic process~\eqref{eq:dv} when either the friction coefficient or the noise intensity gives rise to the cutoff. The case $k(v) = k_0(v)$ and $g(v)=g_0(v)$ will be also referred to as the unlimited L\'evy random walks. 

As intended, in the two cases the stationary distribution function~\eqref{Pst} is of the same form
\begin{equation}\label{IPst}
  P_c^\text{st}(v) = \frac{C_c }{\left(1+v^2\right)^{\big(1-\tfrac1{v_c^2}\big)\tfrac{(1+\alpha)}2}}\, e^{-\tfrac{(1+\alpha)}{2} \left(\tfrac{v}{v_c}\right)^2} 
\end{equation}
with the normalization constant given by the expression
\begin{equation} \label{IPstC}
  C_c= \left[\sqrt\pi U\left(\frac12,\frac{2-\alpha}2 +\frac{(1+\alpha)}{2v_c^2}, \frac{(1+\alpha)}{2v_c^2} \right)\right]^{-1} \,,
\end{equation}
where $U(\cdot,\cdot,\cdot)$ is the confluent hypergeometric function of the second kind. Distribution~\eqref{IPst} can be regarded as a generalized Cauchy distribution with cutoff whose characteristics are studied in detail in Ref.~\cite{Konno2011}.  

\section{Results of numerical simulation}

To analyze the statistical properties of random walks generated by model~\eqref{eq:dv} it has been solved numerically using the SRI2W1 algorithm of the Runge-Kutta methords for strong approximation of stochastic differential equations of the It\^o type with scalar noise \cite{Roussler}. This algorithm has the deterministic order 3.0 and the stochastic order 1.5. Mersenne Twister algorithm by Matsumoto and Nishimura \cite{MTAlg} implemented in GNU Scientific Library 1.14 \cite{GSL} was used in generating random numbers. In integration the time step was set equal to $dt=0.01$, which gives stable results with respect to a decrease or increase of the time step by several times. The integration time was chosen to be equal to $T=10^8$ in order to accumulate enough statistics for analyzing the distribution tails based on individual trajectories, which enables us to avoid the problem of possible partial ergodicity of systems under consideration \cite{me3}. The created computer program was verified, in particular, by reproducing the velocity distribution function~\eqref{IPst} numerically.     

The obtained numerical data were used in order to analyze two characteristics. The first one is the geometric mean $\overline{x}_g(t)$ of the particle displacement $\delta_t x$ during the time interval $t$; it is defined via the formula
\begin{equation}\label{num:1}
   \ln\left[\overline{x}_g(t)\right] = \left<\ln\left(|\delta_t x|\right) \right>\,.
\end{equation}
Here the angle brackets $\langle\ldots\rangle$ denote the time averaging, i.e., partitioning of a generated trajectory of length $T\gg t$ into fragments of duration $t$ and then averaging over the obtained fragments. In the given model for the unlimited L\'evy random walks, i.e., for $v_c=\infty$ the asymptotics of the geometric mean $\overline{x}_g(t)$ as $t\to\infty$ can be written in the form  \cite{me2,me3}       
\begin{equation}\label{num:2}
   \overline{x}_g^L(t) = \Lambda_g t^{1/\alpha}\,,
\end{equation}
where the coefficient
\begin{equation}\label{num:3}
   \Lambda_g = \left[\frac{\Gamma\left(\dfrac{2-\alpha}{2}\right)}{2^{\alpha-1}\Gamma\left(\dfrac{\alpha}{2}\right)\Gamma(\alpha)}\right]^{\tfrac1\alpha}
    \exp\left[\gamma\frac{(1-\alpha)}{\alpha}\right]\,,
\end{equation}
$\Gamma(\ldots)$ is the gamma function, and $\gamma\approx 0.5772$ is the Euler-Mascheroni constant.

The second characteristics is the probability density $P(\rho,t)$ of the particle displacement $\delta_t x$ during the time interval $t$ measured in units of the corresponding geometric mean, i.e., the random variable $\rho = \delta_t x/\overline{x}_g(t)$. Again this distribution density was constructed based on partitioning of a single sufficiently long trajectory. For the unlimited L\'evy random walks the asymptotics of the probability density is known, namely, it is of the form \cite{me3}
\begin{equation}\label{num:4}
   P_L(\rho) =  \frac{2}{\pi}\sin\left(\frac{\pi\alpha}{2}\right)\Gamma(1+\alpha)e^{\gamma(\alpha-1)}\frac1{\rho^{1+\alpha}}  
\end{equation}
as $\rho\to\infty$ and does not depend on the time scale for $t\gg1$.  

\begin{figure}
 \begin{center}
   \includegraphics[width=0.95\columnwidth]{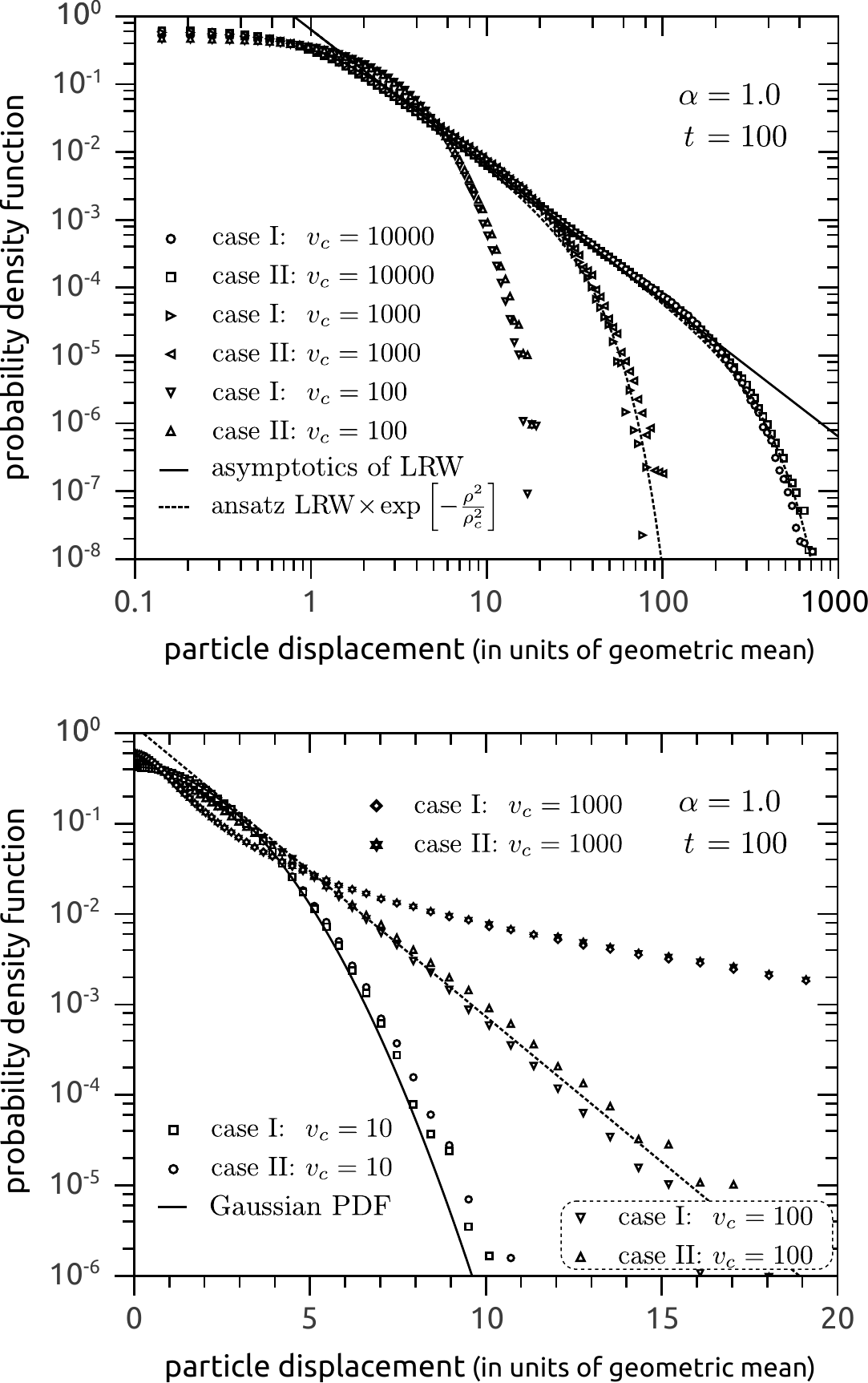}
\end{center}
\caption{Probability density $P(\rho,t)$ as a function of the particle displacement normalized to the the corresponding geometric mean, $\rho(t) =\delta_t x/\overline{x}_g(t)$, for various values of the cutoff velocity $v_c$. The used values of the parameter $\alpha$ and the duration $t$ of the analyzed trajectory fragments are shown inside the frames.} 
\label{F2}
\end{figure}

Figure~\ref{F2} shows the obtained results for the probability density $P(\rho,t)$. Appealing to this figure we can state the following. When the cutoff region is rather distant from the origin in the velocity space, i.e., the parameter $v_c$ is rather large, model~\eqref{eq:dv} descries random walks for which the particle displacement $\delta_t x$ obeys the generalized Cauchy statistics with a certain cutoff. The solid line in the upper frame of Fig.~\ref{F2} visualizes asymptotic~\eqref{num:4} corresponding to the unlimited L\'evy random walks. In this frame the dotted lines display the phenomenological Ansatz 
\begin{equation}\label{num:5}
   P(\rho,t) \sim  P_L(\rho)\exp\left[-\frac{\rho^2}{\rho_c^2}\right] \quad\text{as $\rho\to\infty$} \,,
\end{equation}
where $\rho_{c} \approx 3.2\times10^2$ and $\rho_{c} \approx 3.3\times10$ for $v_c = 10^5$ and $v_c = 10^3$, respectively. The found magnitudes of the parameter $\rho_c$ validate the following construction of the relationship between the critical velocity $v_c$ and the characteristic scale $\Delta_c$ determining the cutoff region $\delta_tx\gtrsim \Delta_c$ in the space of the particle displacement $\{\delta_t x\}$.

As it is known \cite{me1,me2}, for the unlimited L\'evy random walks there is a relation between the particle displacement $\delta_t x$ attained within a given trajectory fragment of duration $t\gg1$ and the extreme fluctuations in the particle velocity within this fragment. Namely, the magnitude of $\delta_t x$ is mainly determined by motion of the particle during the peak of the corresponding time pattern $\{v(t')\}$ with the maximal amplitude $v^\text{max}_t$. Since the characteristic time scale of such extreme velocity fluctuations is about unity (in dimensionless units), the proportionality $\delta_t x\sim c v^\text{max}_t$ with $c\sim 1$ holds. Moreover, the estimate 
\begin{equation}
 c =  \left[ 
      \frac{2\sin\left(\dfrac{\pi\alpha}{2}\right)\Gamma\left(\dfrac{2-\alpha}{2}\right)}
      {\sqrt\pi \alpha\Gamma\left(\dfrac{\alpha+1}{2}\right)}
      \right]^{\tfrac1\alpha}
\end{equation}
can be employed \cite{me1} and for $\alpha =1$ we have $c = 2$. In particular, the geometric mean $\overline{x}_g(t)$ actually evaluates the characteristic amplitude of these extreme peaks gained by the particle velocity during the time interval $t$, i.e., $c\langle v^\text{max}_t\rangle \sim \overline{x}_g(t)$ \cite{me2,me3}. 

The cutoff effect in the statistics of the particle displacement has to become pronounced when the amplitude of extreme velocity fluctuations gets the cutoff region in the velocity space, $v^\text{max}_t\sim  v_c$. It immediately enables us to write the desired relationship
\begin{subequations}\label{fin:1}
\begin{align}
\label{fin:1a}
   \Delta_c & \sim c v_c\,,
\\
\intertext{in the used notations it becomes}
\label{fin:1b}
    \rho_c(t)\cdot \overline{x}_g(t) & \sim c v_c \,.
\end{align}
\end{subequations}
According to the results to be presented below as well as obtained in Ref.~\cite{me3}, the estimate $\overline{x}_g(t)\approx 100$ holds for the unlimited L\'evy random walks when the parameter $\alpha = 1$ and the time scale $t = 100$. The found magnitudes for the quantity $\rho_c$ and the geometric mean $\overline{x}_g$ together with the values of the parameter $v_c$ used in the numerical simulations fit estimate~\eqref{fin:1b} very well, justifying the given arguments. 

As the time scale $t$ increases the intermediate region matching the L\'evy type asymptotics~\eqref{num:4} should shrink and disappear completely when the equality $\overline{x}_g(t)\sim v_c$ if achieved. This conclusion is justified by the obtained numerical data illustrated in Fig.~\ref{F2}. It plots the probability density functions of the particle displacement that were constructed for the fix time scale $t= 100$ but different values of the parameter $v_c$. Naturally, for $\overline{x}_g(t)\gg v_c$ the probability density function $P(\rho)$ must be of the Gaussian form, which also is demonstrated in Fig.~\ref{F2}, the lower fragment. It should be pointed out that the crossover from the generalized Cauchy distribution with cutoff to the Gaussian distribution cannot be represented as a continuous transformation of Ansatz~\eqref{num:5}. In fact, the probability density $P(\rho)$ for the intermediate values of the parameter  $v_c =100\approx \overline{x}_g(t)$ for $\alpha =1$ and $t = 100$ looks like an exponential function. To make it clear the Ansatz
\begin{equation*}
   P(\rho) = 1.2 e^{-0.74\rho}  
\end{equation*}
is depicted by the dashed line in Fig.~\ref{F2}, the lower fragment.
  
In addition, Figure~\ref{F2} demonstrates the fact that statistics of the particle displacement depends rather weakly on the individual details of the kinetic coefficients $k(v)$ and $g(v)$, only their ratio $k(v)/g^2(v)$ matters to it. It could be explained if the statistics of the extreme velocity fluctuations is determined mainly by the stationary distribution function of the particle velocity, which is worthy of individual investigation.

\begin{figure}
 \begin{center}
   \includegraphics[width=0.95\columnwidth]{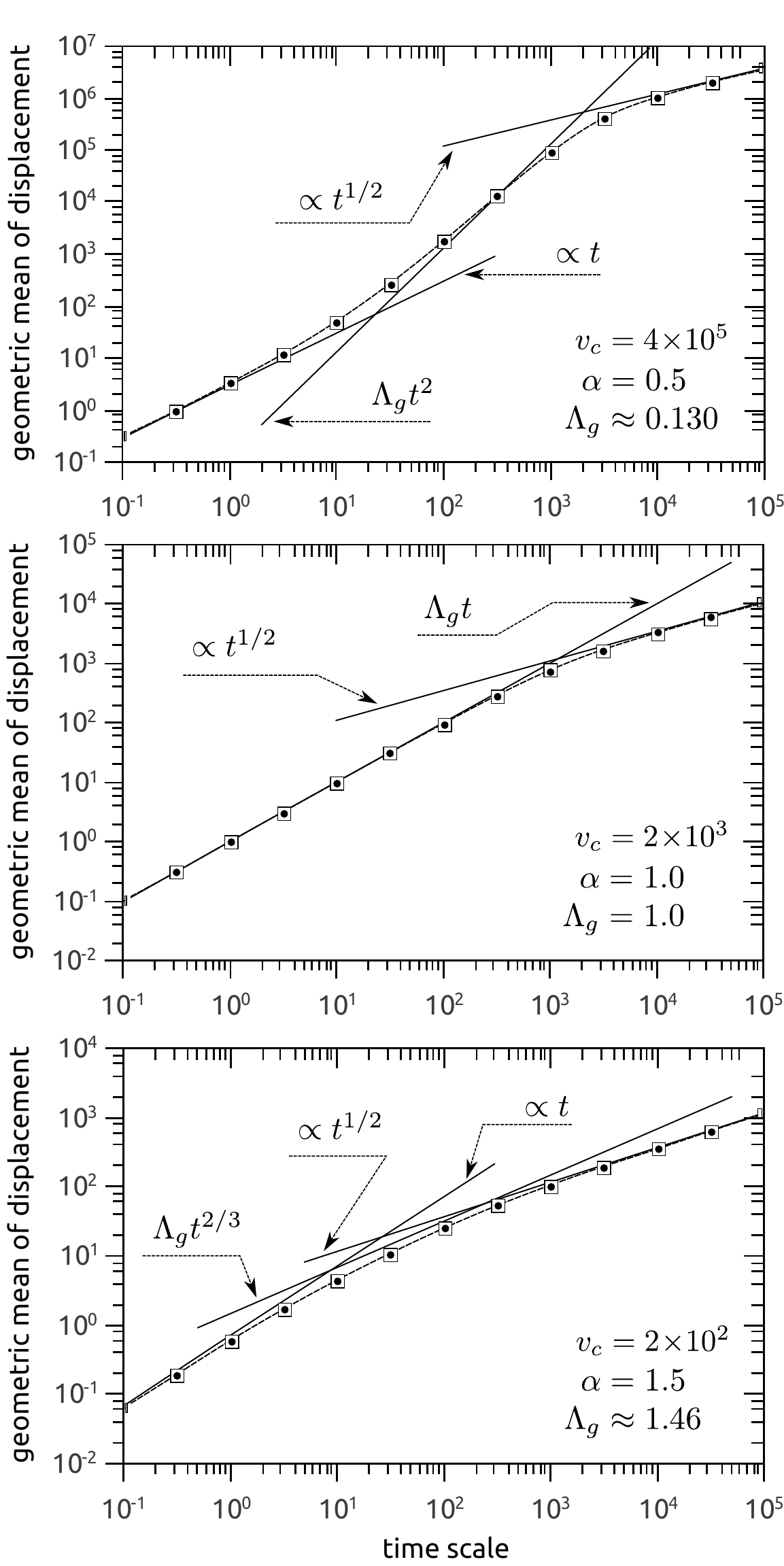}
\end{center}
\caption{The geometric mean of the particle displacements vs the duration of time steps in the partitioning of the random walks. The symbols $\Box$ and \textbullet\ depict the data points for cases I and II, respectively.}
\label{F3}
\end{figure}

Figure~\ref{F3} visualizes the time dependence of the geometric mean $\overline{x}_g(t)$ for several values of the parameter $\alpha$, namely, $\alpha = 0.5$, 1.0, and 1.5. These values match the superballistic, quasiballistic, and superdiffusive type of the expected L\'evy scaling law $\overline{x}_g(t)\propto t^{1/\alpha}$. The choice of the corresponding values of the cutoff velocity $v_c$ will be explained below. 

Appealing to the data plotted in Fig.~\ref{F3} we can single out three characteristic stages in the  $\overline{x}_g(t)$-dependence. First, on small scales the time dependence $\overline{x}_g(t)$ is found to be linear, as it must because of the strong velocity correlations for $t\lesssim1$. Second, there is a certain intermediate stage that can be classified as the region of the L\'evy scaling law. In fact, according to Ref.~\cite{me3} random walks governed by model~\eqref{eq:dv} with $v_c = \infty$ exhibit the L\'evy type scaling law~\eqref{num:2}, i.e.,  $\overline{x}^L_{g}(t) = \Lambda_g t^{1/\alpha}$,  when the duration of the trajectory partitioning fragments exceeds a certain value about 100, i.e., $t\gtrsim100$ (in dimensionless units). So to make the intermediate L\'evy asymptotics feasible, in the numerical simulation the critical velocity $v_c$ was chosen to exceed the geometric mean $\overline{x}_{g}(t)|_{t=100}$ for unlimited L\'evy random walks tenfold. The employed values $v_c$ meeting this requirement are noted in Fig.~\ref{F3}. In order to clarify the fact that in the cases under consideration the geometric mean $\overline{x}_{g}(t)$ does exhibit the intermediate L\'evy asymptotics the function $\Lambda_gt^{1/\alpha}$ is also plotted in Fig.~\ref{F3}. The lines visualizing this function are seen to be practically tangents to the curves interpolating the numerical data. Finally, for large values of time scale the cutoff effect becomes substantial and, as it must, the region of the classical asymptotics $\overline{x}_{g}(t)\propto t^{1/2}$ of Brownian motion arises. 

It should be pointed out that in both cases I and II actually the same geometric mean $\overline{x}_{g}$ was obtained, which again justifies the previously drawn conclusion about the weak dependence of the given truncated L\'evy random walks on the individual details of the kinetic coefficients $k(v)$ and $g(v)$.

\section{Conclusion}

The paper is devote to the development of a continuous theory of L\'evy random walks that was recently initiated in Refs.~\cite{me1,me2,me3}. Namely, a continuous Markovian model for truncated L\'evy random walks has been developed. It is based on the nonlinear stochastic differential equation~\eqref{eq:dv} governing the velocity $v$ of a wondering particle. The cutoff effects in the distribution of the particle velocity and displacement during a certain time interval can be caused by the velocity dependence of the friction coefficient as well as the saturation in the growth of the Langevin force intensity with the particle velocity. Both the mechanisms are studied individually and to be able to compare them against  each other the system parameters have been chosen in such a manner that the velocity distribution be the same in both the cases. The one-dimensional system was analyzed numerically employing a strong stochastic Runge-Kutta methor of the stochastic order 1.5 \cite{Roussler}. The obtained numerical data were used to calculate the geometric mean $\overline{x}_g(t)$ of the particle displacement $\delta_t x$ during the time interval $t$ and the probability density function $P(\rho,t)$ of this displacement normalized to its geometric mean, $\rho = \delta_t x/\overline{x}_g(t)$.

It has been demonstrated that the two mechanisms do give rise to truncated L\'evy random walks on time scales exceeding substantially the ``microscopic'' time characterizing strong correlations in the velocity fluctuations. It is the case when the region $v\gtrsim v_c$, wherein the friction coefficient nonlinearity as well as the noise intensity saturation become substantial, is rather distant from the origin in the velocity space. As far as particular results are concerned, the following should be noted. 

First, the time dependence of the geometric mean $\overline{x}_g(t)$ has been found to contain three characteristic asymptotics following one another as the time scale $t$ grows. The initial one matching the ballistic regime $\overline{x}_g(t)\propto t$ takes place for $t\lesssim1$. The intermediate L\'evy scaling law  $\overline{x}_g(t)\propto t^{1/\alpha}$, which can represent the superballistic, quasiballistic, or superdiffusive regimes depending on the parameter $\alpha\in(0,2)$, is implemented for the time scales $t\gtrsim100$. On larger scales the cutoff effects become substantial and the $\overline{x}_g(t)$-dependence converts into the classical law of Brownian motion, $\overline{x}_g(t)\propto t^{1/2}$. 

Second, in the region of the L\'evy scaling law the constructed probability density of the particle displacement  $\delta_t x$ has been shown to be of the generalized Cauchy form with an exponential cutoff pronounced in the region $\delta_tx\gtrsim \Delta_c$. Using the numerical data the relationship $\Delta_c\sim v_c$ between the cutoff scale in the space of particle displacement and the velocity $v_c$ characterizing the nonlinearity of the kinetic coefficients is justified.                      

Finally, the properties of the given truncated L\'evy random walks have turned out to depend weakly on the individual characteristics of the friction coefficient $k(v)$ and the noise intensity $g(v)$. Only the ratio $k(v)/g^2(v)$ specifying directly the stationary distribution of the particle velocity matters to them.

\end{document}